\documentclass{aip-cp}

\pdfoutput=1
\usepackage[numbers]{natbib}
\usepackage{graphicx}
\usepackage[greek,english]{babel}
\usepackage[bf]{caption}
\addto\captionsenglish{\renewcommand{\figurename}{FIGURE}}
\begin{document}

\title{Precise Charged Particle Timing with the PICOSEC Detector}

\author[aff2]{J. Bortfeldt}
\author[aff2]{F. Brunbauer}
\author[aff2]{C. David}
\author[aff3]{D. Desforge}
\author[aff5]{G. Fanourakis}
\author[aff2]{J. Franchi}
\author[aff7]{M. Gallinaro}
\author[aff11]{F. Garc\'{i}a}
\author[aff3]{I. Giomataris}
\author[aff9]{T. Gustavsson}
\author[aff3]{C. Guyot}
\author[aff3]{F.J. Iguaz}
\author[aff3]{M. Kebbiri}
\author[aff1]{K. Kordas}
\author[aff3]{P. Legou}
\author[aff4]{J. Liu}
\author[aff2]{M. Lupberger}
\author[aff3]{O. Maillard}
\author[aff1]{\mbox{I. Manthos}\corref{cor1}}
\author[aff2]{H. M\"{u}ller}
\author[aff1]{V. Niaouris}
\author[aff2]{E. Oliveri}
\author[aff3]{T. Papaevangelou}
\author[aff1]{\mbox{K. Paraschou}}
\author[aff10]{M. Pomorski}
\author[aff4]{B. Qi}
\author[aff2]{F. Resnati}
\author[aff2]{L. Ropelewski}
\author[aff1]{\mbox{D. Sampsonidis}}
\author[aff2]{T. Schneider}
\author[aff3]{P. Schwemling}
\author[aff3]{L. Sohl}
\author[aff2]{M. van Stenis}
\author[aff2]{\mbox{P. Thuiner}}
\author[aff6]{Y. Tsipolitis}
\author[aff1]{S.E. Tzamarias}
\author[aff8,aff12,aff13]{R. Veenhof}
\author[aff4]{X. Wang}
\author[aff2]{\mbox{S. White}}
\author[aff4]{Z. Zhang}
\author[aff4]{Y. Zhou}

\affil[aff3]{IRFU, CEA, Universit´e Paris-Saclay, F-91191 Gif-sur-Yvette, France}
\affil[aff2]{European Organization for Nuclear Research (CERN), CH-1211 Geneve 23, Switzerland}
\affil[aff4]{State Key Laboratory of Particle Detection and Electronics, University of Science and Technology of China, Hefei CN-230026, China}
\affil[aff1]{Department of Physics, Aristotle University of Thessaloniki, University Campus, GR-54124, Thessaloniki, Greece.}
\affil[aff5]{Institute of Nuclear and Particle Physics, NCSR Demokritos, GR-15341 Agia Paraskevi, Attiki, Greece}
\affil[aff6]{National Technical University of Athens, Athens, Greece}
\affil[aff7]{Laborat\'{o}rio de Instrumentac\~{a}o e F\'{i}sica Experimental de Part\'{i}culas, Lisbon, Portugal}
\affil[aff8]{RD51 collaboration, European Organization for Nuclear Research (CERN), CH-1211 Geneve 23, Switzerland}
\affil[aff9]{LIDYL, CEA, CNRS, Universit Paris-Saclay, F-91191 Gif-sur-Yvette, France}
\affil[aff10]{CEA-LIST, Diamond Sensors Laboratory, CEA Saclay, F-91191 Gif-sur-Yvette, France}
\affil[aff11]{Helsinki Institute of Physics, University of Helsinki, FI-00014 Helsinki, Finland}
\affil[aff12]{National Research Nuclear University MEPhI, Kashirskoe Highway 31, Moscow, Russia}
\affil[aff13]{Department of Physics, Uludag University, TR-16059 Bursa, Turkey}

\corresp[cor1]{Corresponding author: ioannis.manthos@cern.ch}

\maketitle

\begin{abstract}
The experimental requirements in near future accelerators (e.g. High Luminosity-LHC) has stimulated intense interest in development of detectors with high precision timing capabilities. With this as a goal, a new detection concept called PICOSEC, which is based to a ``two-stage'' MicroMegas detector coupled to a Cherenkov radiator equipped with a photocathode has been developed. Results obtained with this new detector yield a time resolution of 24\,ps for 150\,GeV muons and 76\,ps for single photoelectrons. In this paper we will report on the performance of the PICOSEC in test beams, as well as simulation studies and modelling of its timing characteristics.
\end{abstract}
\section{INTRODUCTION}
It is very well known that in order to take advantage of the High Luminosity (HL) of the Large Hadron Collider (LHC) we need to have the ability to tag events which are promising to carry signatures of new physics. The experimental signature of such events (e.g. forward energetic jets) are difficult to be identified in the very dense topology of the $\sim 140$ interactions per beam crossing in the HL-LHC. The LHC experiments are forced to use precise timing detectors to explore the possibilities the accelerators offer. Today, there are available detectors that provide timing of a few ps per Minimum Ionising Particle (MIP), par example the Micro Channel Plate (MCP) detector. However, since the hermetic approach at the LHC and future accelerator experiments requires large area coverage, it is only natural to investigate more economical solutions, such as Micro-Patern Gas and Silicon structures, to offer such timing capabilities. However, since the necessary time resolution for pileup mitigation is of the order of 20-30\,ps, both technologies require significant modifications to reach the desired performance. Eventually, there is need for large area detectors, resistant to radiation damage with $\sim 10\,\mathrm{ps}$ timing capabilities, that will also find applications in various domains beyond Particle Physics experiments. Photon's energy/speed measurements and correlations for Cosmology, optical tracking for charged particles, 4D tracking in the future accelerators etc. In this paper it is reported the performance of the very successful technology of MicroMegas (MM) detectors, for precise timing purposes. Focus is given to the test beam calibration runs, data analysis and the physics of this detector.
\section{THE PICOSEC DETECTOR CONCEPT}
\begin{figure}
\centering
\begin{minipage}{.5\textwidth}
\centering
\includegraphics[width=1.\textwidth]{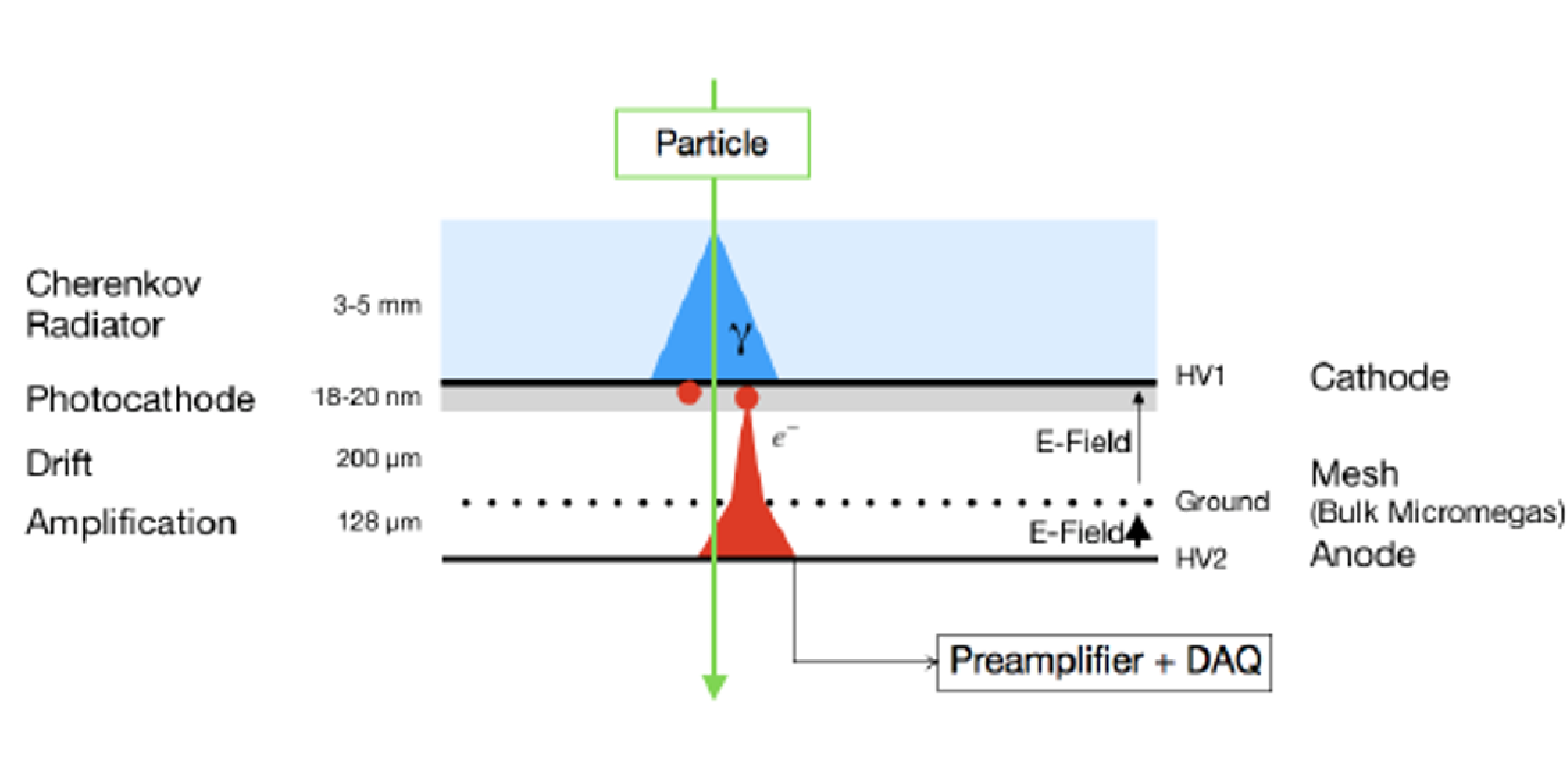}
\end{minipage}
\begin{minipage}{.5\textwidth}
\centering
\includegraphics[width=1.\textwidth]{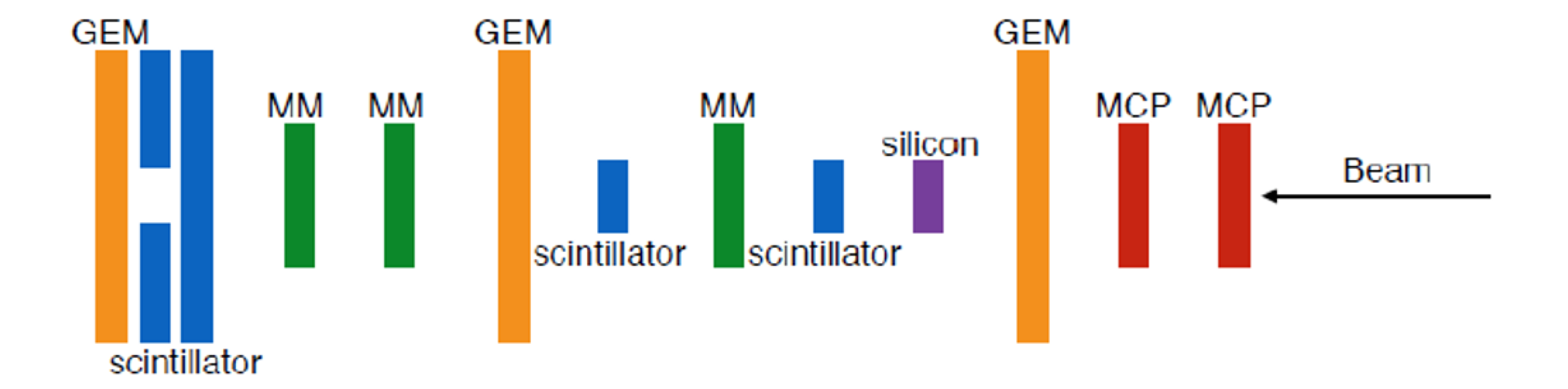}
\end{minipage}
\caption{The layout of the PICOSEC MicroMegas detector \citep{pico24} (left). Sketch of the experimental setup in the 150\,GeV muon test beam (right).}
\label{fig:fig1}
\end{figure}

There is a seminal paper by Sauli \citep{sauli} that explicitly proves that proportional gaseous detectors, due to the statistics of the ionization can hardly exceed the ns-level for the time resolution per particle ($\sim 6\,\mathrm{ns}$ for typical chamber dimensions). Recently, the RD-51 PICOSEC collaboration developed a detection technique \citep{papaevangelou} and published work \citep{pico24, iguaz} demonstrating that a special MicroMegas detector can reach a timing resolution better than 25\,ps. In this detector (hereafter PICOSEC-MM) the drift region has been reduced to less than $200\,\textrm{\selectlanguage{greek}m\selectlanguage{english}m}$ in order to practically minimize the possibility of  ionisation in the drift region, whilst the anode region remains at the typical size of MM ($128\,\textrm{\selectlanguage{greek}m\selectlanguage{english}m}$). A primary particle, before entering the drift region, passes through a 3\,mm Cherenkov radiator of $MgF_{2}$ and the produced photons generate photoelectrons on a photocathode, placed just below the radiator, in contact with the gas volume of the drift region, as on Fig. \ref{fig:fig1} (left). The photocathode has been deposited on a thin film of semitransparent Cr which is used to provide conductivity to the cathode. The photoelectrons are produced almost synchronously and start avalanches in the drift region due to the high electric field produced by modest voltage differences (of the order of $\sim 400\,\textrm{V}$). The preamplification avalanche in the drift region is shown schematically in Fig. \ref{fig:fig1} (left).
\subsection{Muon Test Beam Results}  
\begin{figure}
\centering
\begin{minipage}{.33\textwidth}
\centering
\includegraphics[width=1.\textwidth]{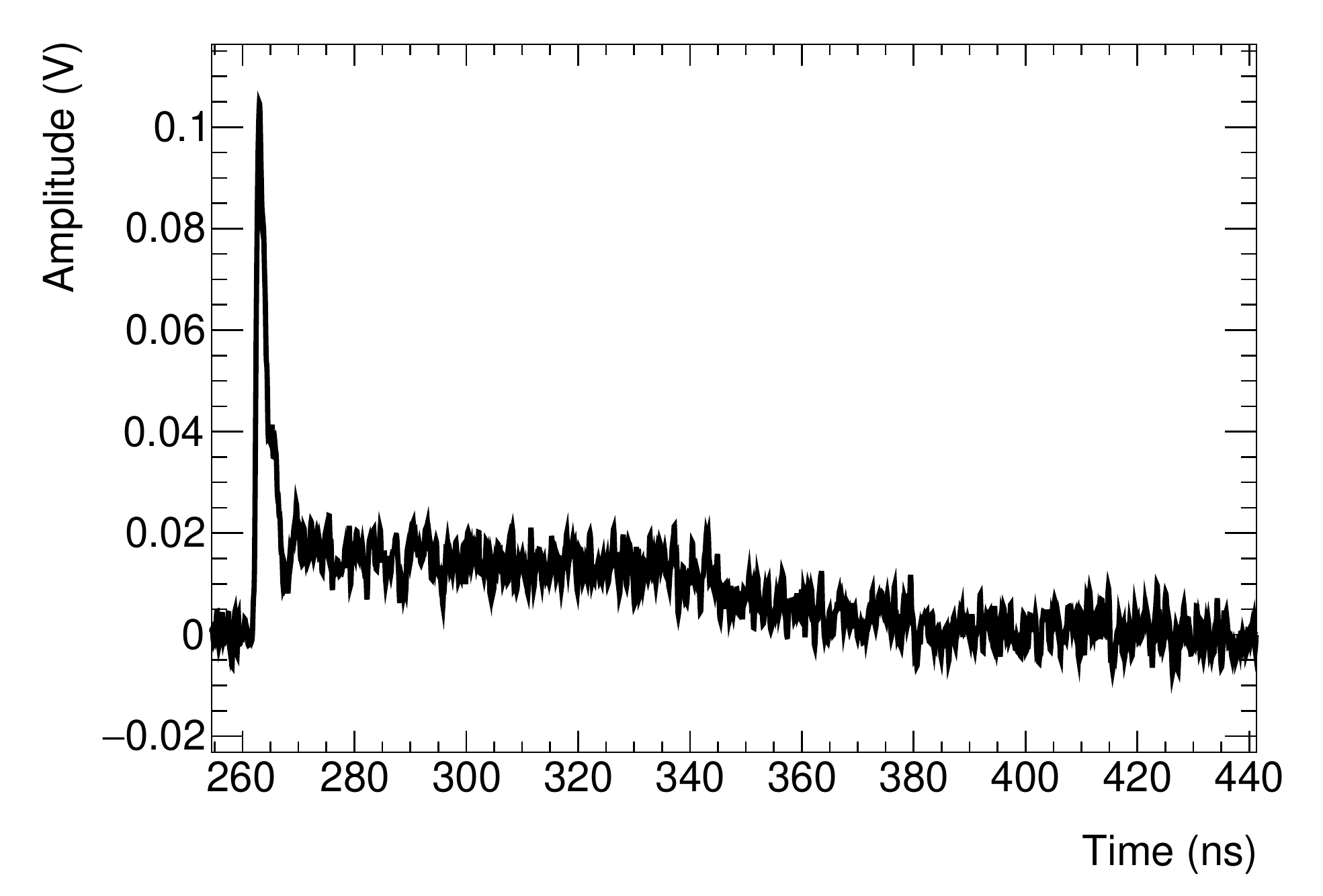}
\end{minipage}
\begin{minipage}{.33\textwidth}
\centering
\includegraphics[width=.8\textwidth]{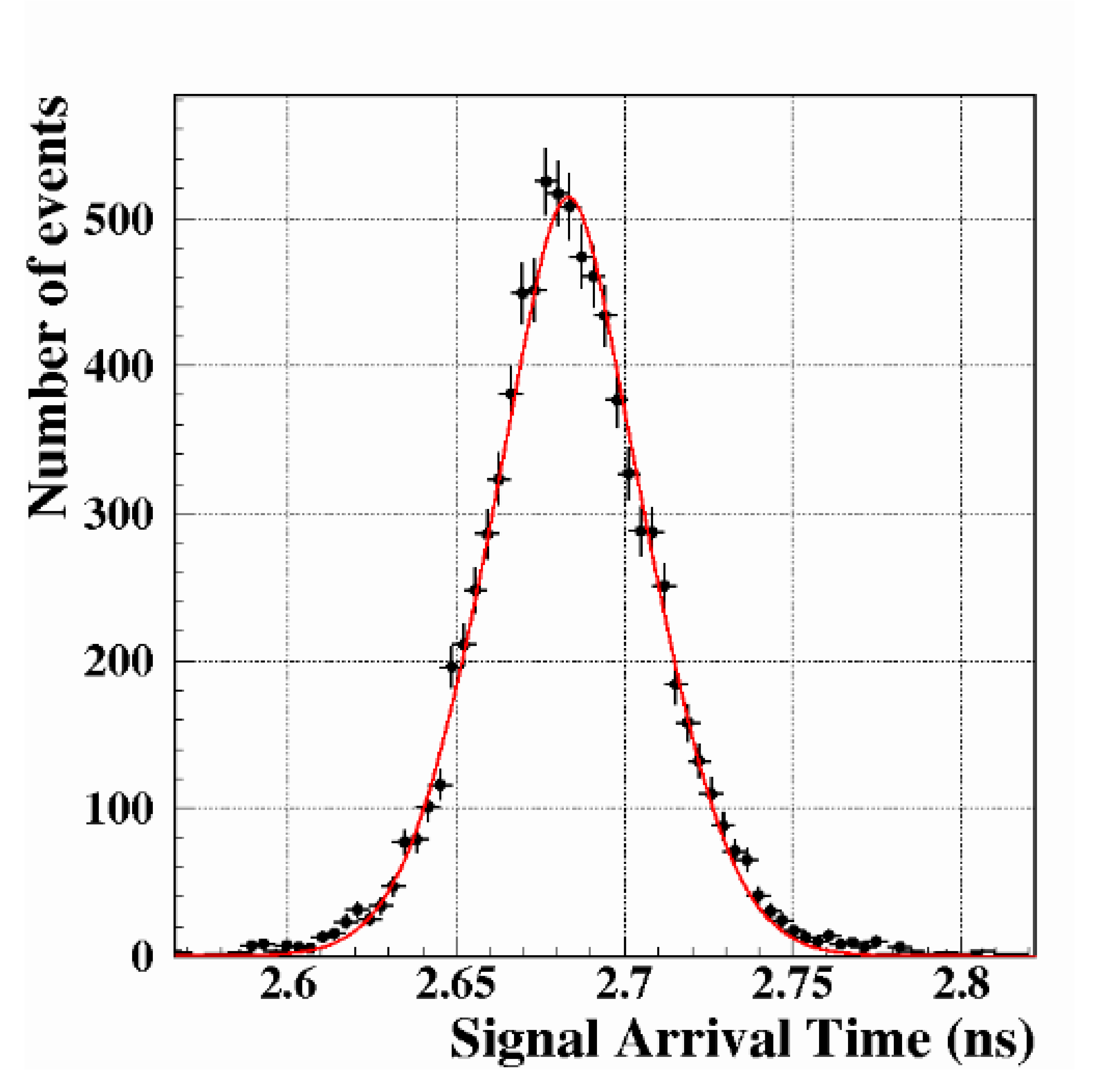}
\end{minipage}
\begin{minipage}{.33\textwidth}
\centering
\includegraphics[width=.75\textwidth]{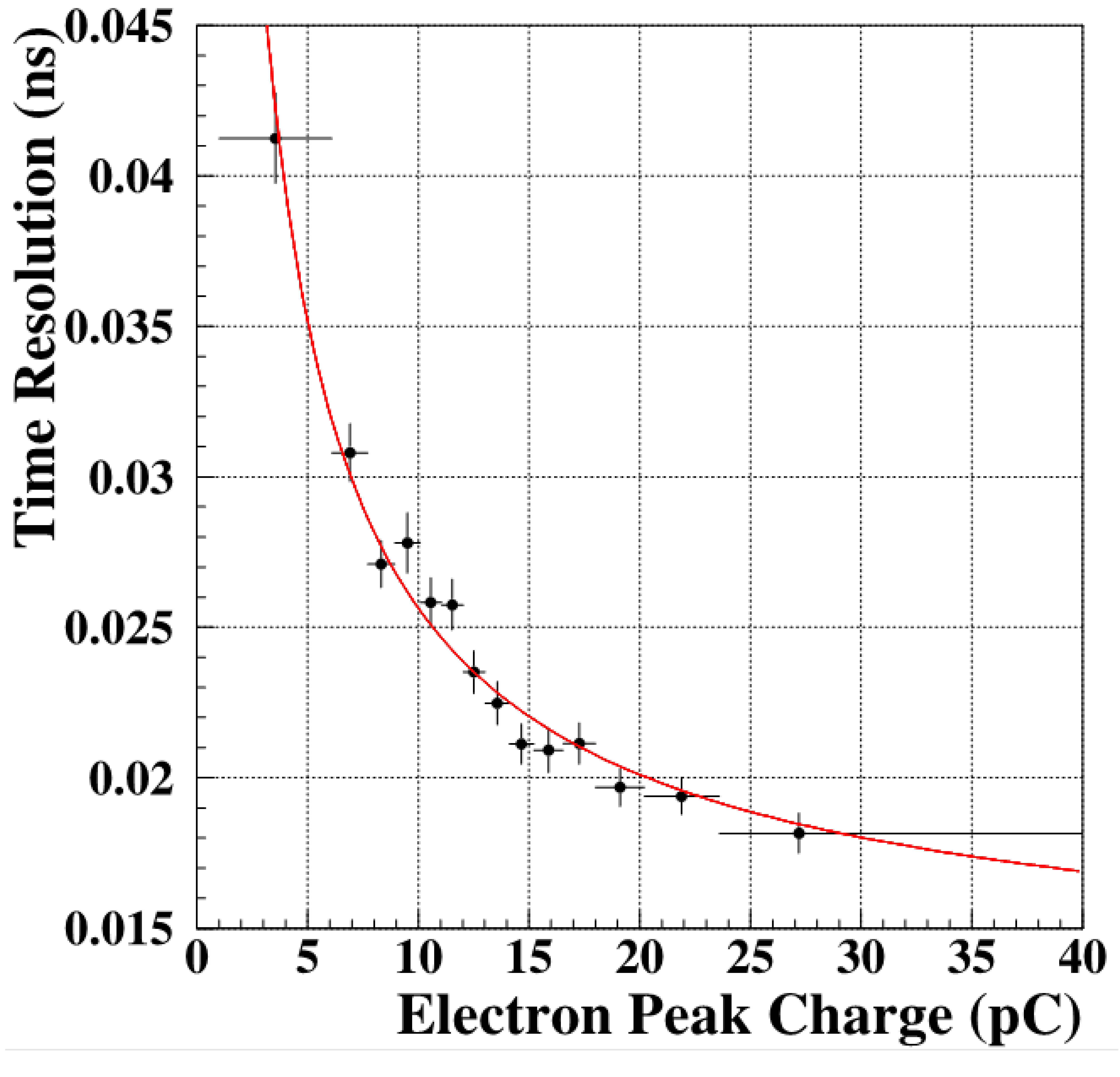}
\end{minipage}
\caption{A typical pulse produced by the PICOSEC-MM detector (left). SAT distribution for 150\,GeV muons, and the fit with a two Gaussian function resulting to 24\,ps timing resolution, for anode and drift voltage of 275\,V and 475\,V respectively (center). The dependence of the resolution as a function of the e-peak pulse's charge (right).}
\label{fig:fig2}
\end{figure}

\begin{figure}
\centering
\begin{minipage}{.25\textwidth}
\centering
\includegraphics[width=0.95\textwidth]{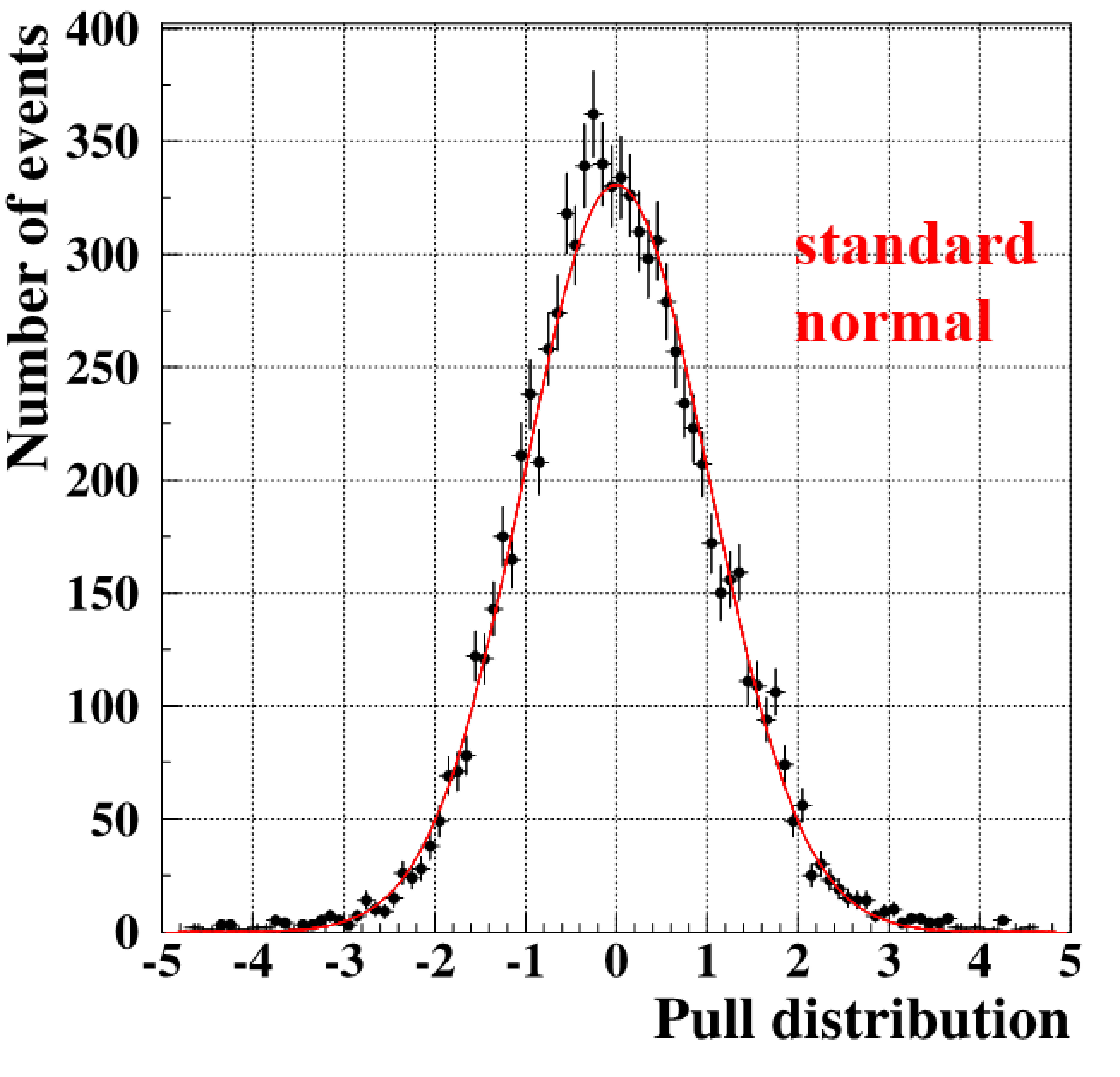}
\end{minipage}
\begin{minipage}{.25\textwidth}
\centering
\includegraphics[width=0.92\textwidth]{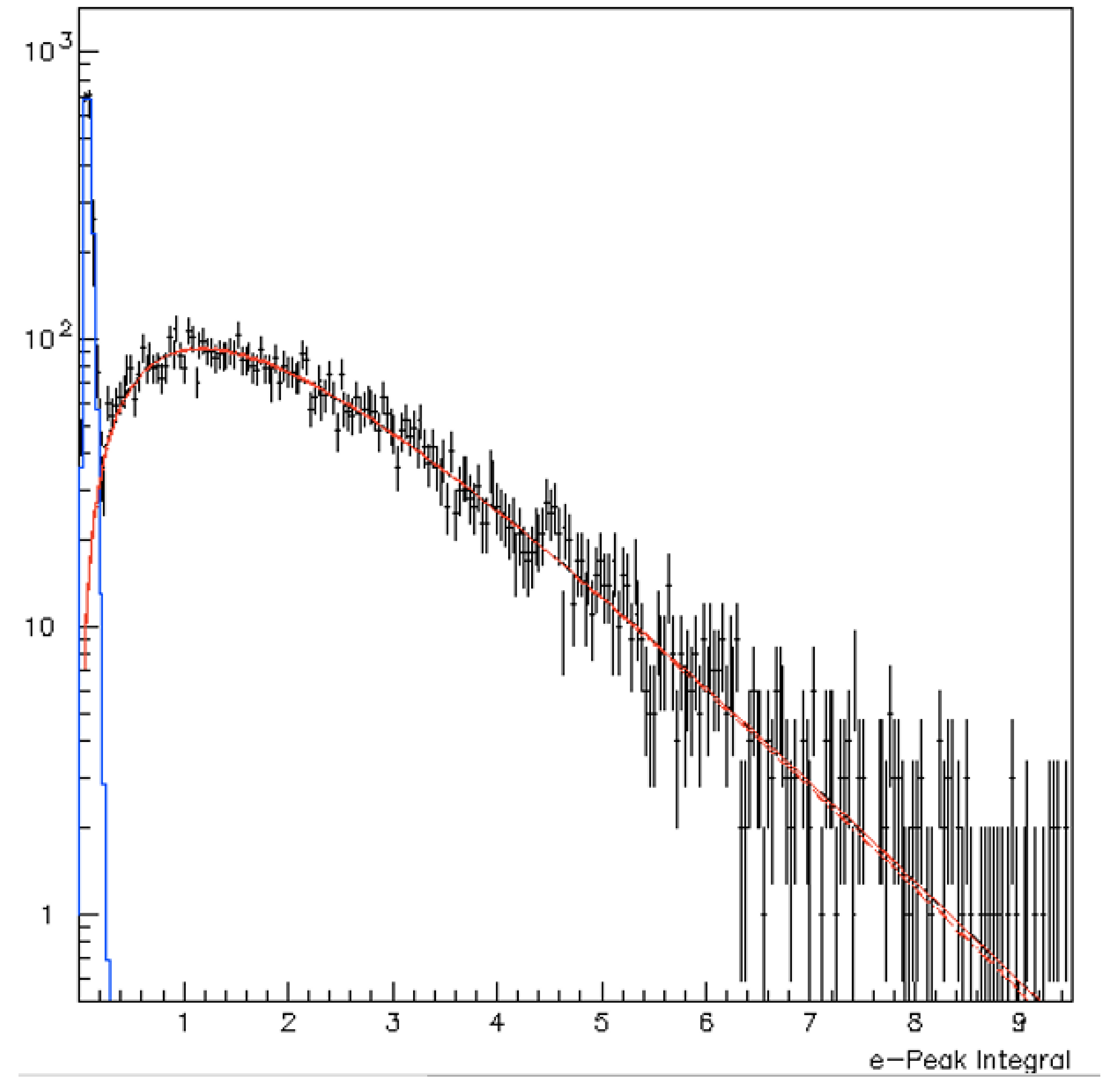}
\end{minipage}
\begin{minipage}{.5\textwidth}
\centering
\includegraphics[width=1.\textwidth]{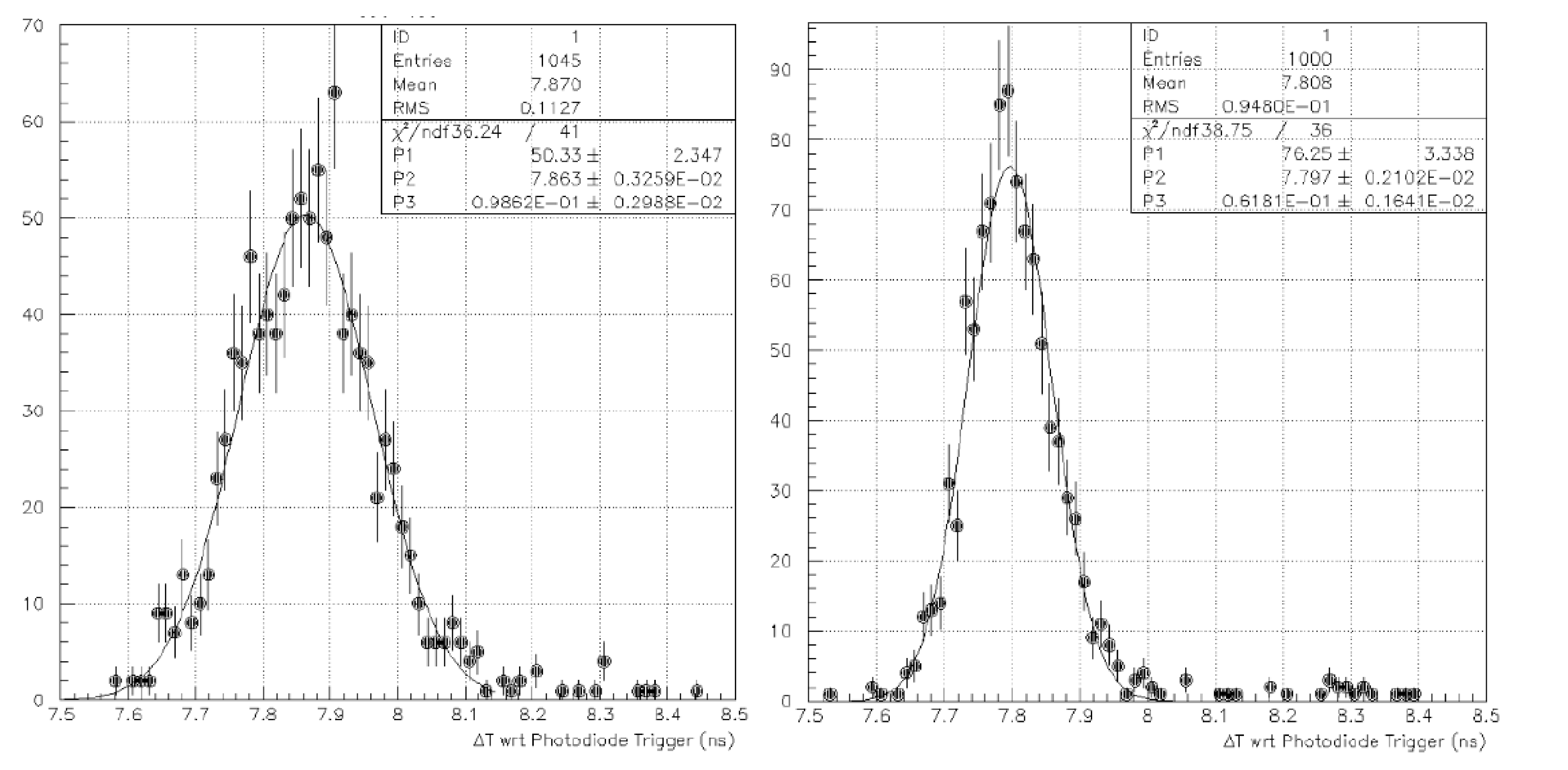}
\end{minipage}
\caption  {The pull distribution of the timing resolution dependence on the e-peak charge on an event by event basis (left). Distribution of the e-peak charge collected by the detector response to a single photoelectron. This distribution was parametrized by a Polya function (red), whilst the noise (blue) is parametrized and rejected (central left). SAT distribution for two e-peak charge regions: The first corresponds to low charge values while the second to high charge values. Mean values are different, as larger pulses are coming earlier and with better timing resolution (central right and right plot respectively).}
\label{fig:fig3}
\end{figure}

The PICOSEC -MM detector has been tested extensively at the CERN SPS H4 secondary beamline. The experimental setup \citep{pico24}, shown in Fig. \ref{fig:fig1} (right) comprise scintillation counters for triggering, GEM detectors for tracking the muons of the beam, MCP detectors to provide time reference with a resolution better than 6\,ps and the under test PICOSEC-MM detectors, with COMPASS gas filling ($Ne + 10\% C_{2}H_{6} + 10\% CF_{4}$) at 1\,bar. The PICOSEC-MM signal was amplified by a CIVIDEC amplifier (2\,GHz, 40\,dB) and the waveform was digitized by a 2.5\,GHz oscilloscope at a rate of 20\,GSamples/s (i.e. one sample every 50\,ps). A typical pulse of the PICOSEC-MM responding to a muon is presented in Fig. \ref{fig:fig2} (left). The very fast component with a rise time of $\sim 500\,\textrm{ps}$ and a duration of the order of ns is due to the induction of current on the anode by the fast moving electrons, whilst the slow component due to the slow moving ions is extended up to several hundreds ns. The collected digitised waveforms were analysed offline. Standard analysis procedure was applied to determine the beginning of the pulse, the electron peak  (hereafter e-peak) amplitude, the end of the e-peak and the arrival time of the pulse. A crucial point in this analysis was to estimate accurately the pulse characteristics, especially the arrival time, even if there is a non-negligible noise contribution to the digitised waveform. It was found that by processing the signal using  filtering algorithms \citep{niaouris} the noise contribution was reduced but the very fast e-peak leading edge was affected. A processing analysis algorithm was developed, by fitting the leading edge of the e-peak using a logistic function, which offers the desired accuracy to estimate the Signal Arrival Time (SAT), which was defined as the point that the fitted leading edge was crossing a threshold of the 20\% of the peak maximum (i.e. the standard technique of the Constant Fraction Discrimination (CDF)).
Results achieved with 150\,GeV muon beam with 275\,V anode and 475\,V drift voltage settings are presented in Fig. \ref{fig:fig2} (center) where the distribution of the SAT with respect to the MCP time reference is shown. The distribution is fitted by the sum of two Gaussians and the combined Root Mean Square (RMS) was found to be $24.0\pm 0.3\,\textrm{ps}$ which is the timing resolution per MIP of the detector on these voltage settings. In the right plot of Fig. \ref{fig:fig2} the RMS versus e-peak charge is shown. These distributions (i.e. SAT distributions in narrow bins of the e-peak charge) where found to be Gaussian, however as shown in the right plot of Fig. \ref{fig:fig2} the PICOSEC-MM resolution (RMS) depends on the e-peak charge. This dependence is also the reason that the distribution at the central plot of Fig. \ref{fig:fig2}, corresponding to a wide spectrum of e-peak charge, deviates from the Gaussian shape. Using the parametrization of the resolution on the e-peak charge on event by event basis, the pull distribution is determined as shown in the left of Fig. \ref{fig:fig3}. The fact that the pull distribution is, in a very good approximation, normal Gaussian, supports the argument described above.
\subsection{Single Photoelectron Test Results}
The PICOSEC-MM signal, generated by a muon, is the sum of the detectors response to each of the produced photoelectrons. To study the PICOSEC-MM timing characteristics, the detector response to a single photoelectron was also studied in special runs in Ref. \citep{pico24}. The photocathode was illuminated using a femptosecond laser beam with very low intensity, so as to produce a single photoelectron. The laser beam was split, a portion was illuminating a precise photodiode, which provides a time reference, while the other part illuminates the PICOSEC-MM. Both signals were digitised by a very fast oscilloscope. The distribution of the e-peak charge corresponding to a single photoelectron is presented in Fig. \ref{fig:fig3} (central left). The experimental distribution has been fitted by a Polya function plus a distribution accounting for the background contribution which has been modelled using out of time events in Ref. \citep{notea}.
\begin{figure}
\centering
\begin{minipage}{.28\textwidth}
\centering
\includegraphics[width=1.\textwidth]{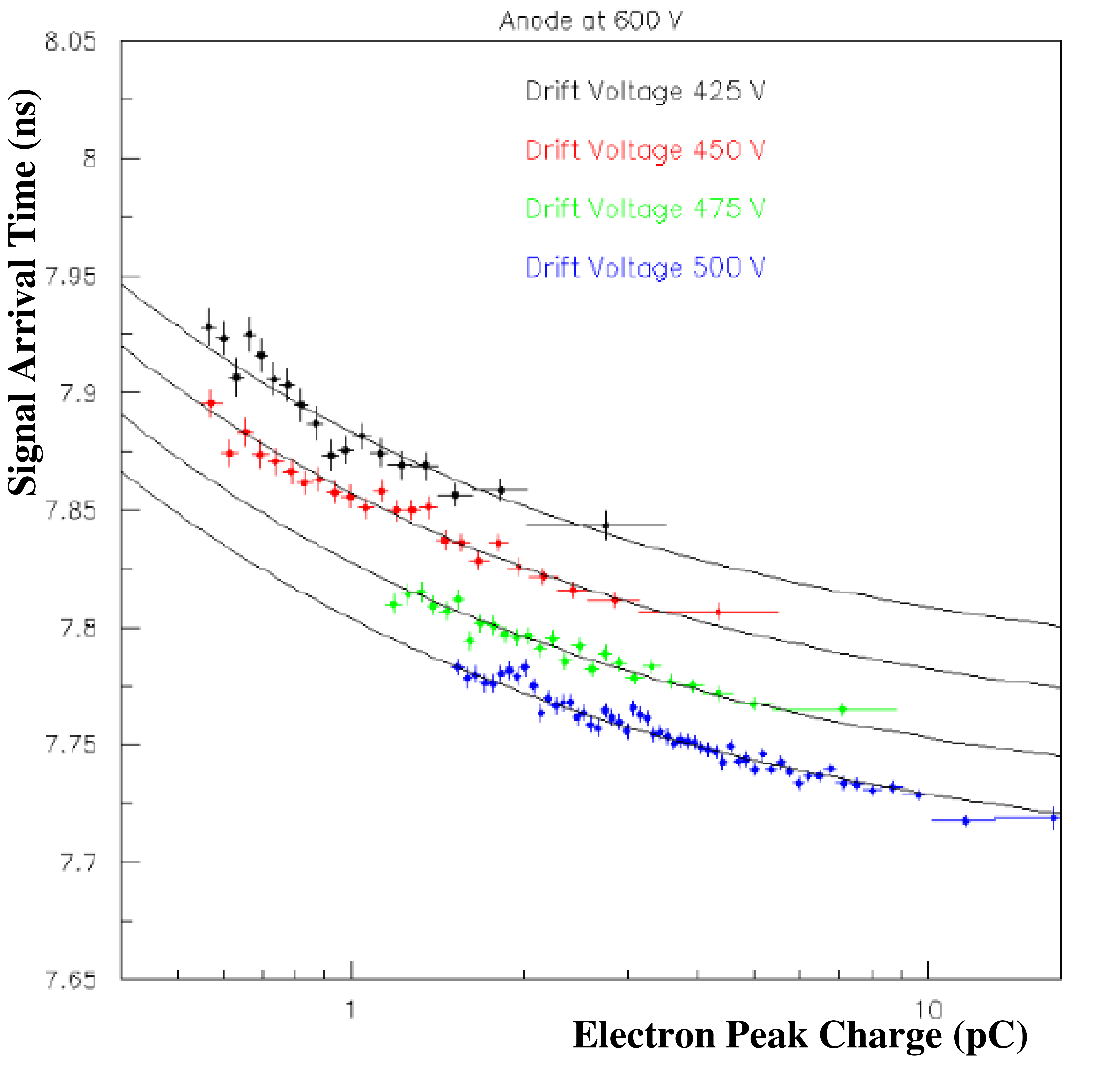}
\end{minipage}
\begin{minipage}{.33\textwidth}
\centering
\includegraphics[width=1.\textwidth]{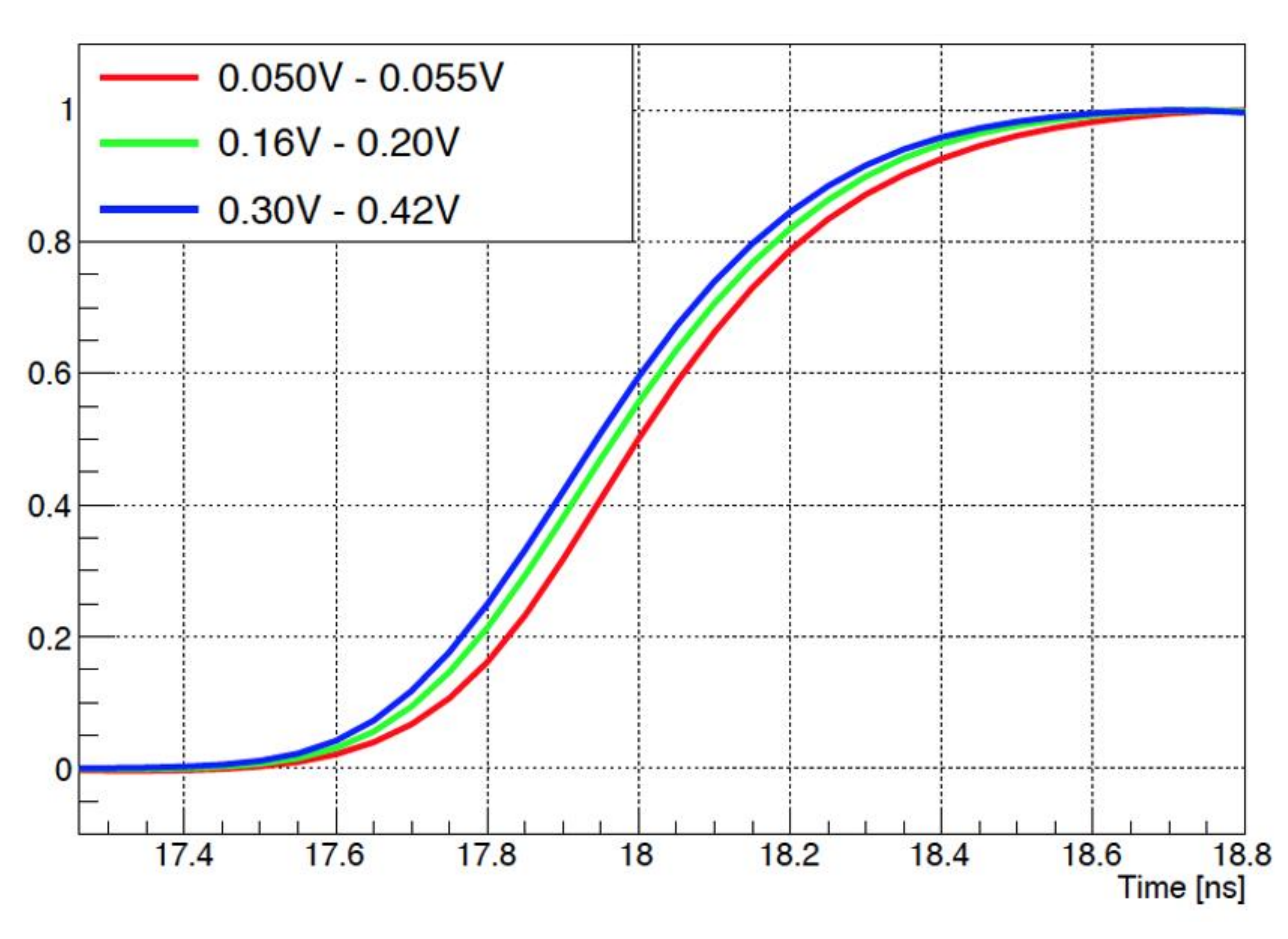}
\end{minipage}
\centering
\begin{minipage}{.28\textwidth}
\centering
\includegraphics[width=1.\textwidth]{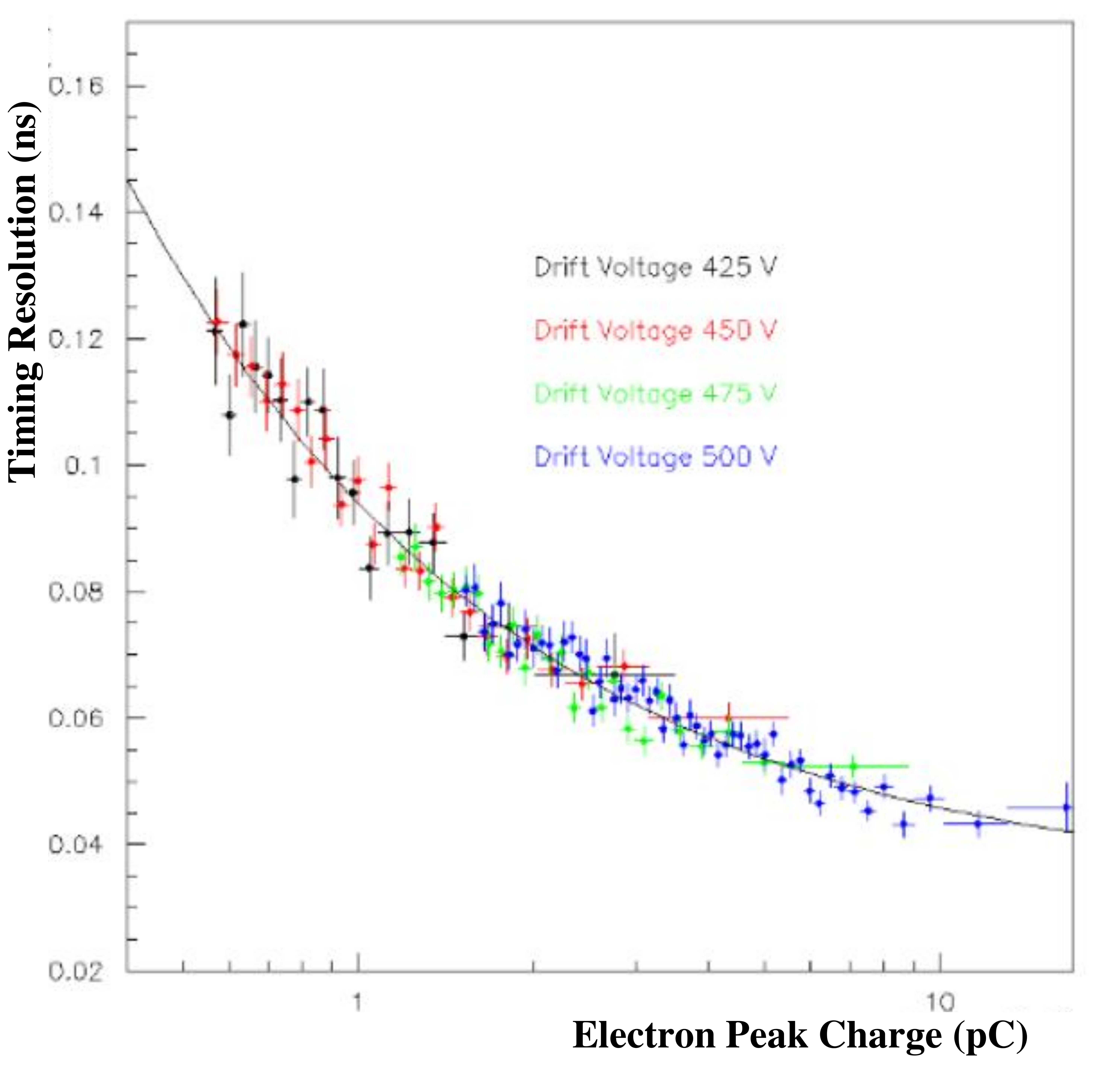}
\end{minipage}
\caption{SAT as a function of the e-peak size for several drift voltages and anode at 600\,V (left). Average of pulses corresponding to different charge regions normalized to the unit. The shape does not change but higher pulses arrive earlier (center). The timing resolution dependence on the e-peak size, for constant anode field 600\,V and different drift voltages as mentioned in the plot (right).}
\label{fig:fig4}
\end{figure}
The SAT distribution (that is the time difference of the PICOSEC-MM pulse arrival time and the reference photodiode pulse timing) for waveforms with e-peak charges in two regions, is shown in center right and right plots of Fig. \ref{fig:fig3}. The first corresponds to low charge values (1.14 - 1.69\,pC) and the second to high charge values (7.96 - 10.40\,pC). As shown in these plots, both the mean values and the RMSs of the two distributions are different, signifying that larger pulses are coming earlier and with better timing resolution than smaller pulses. By studying the SAT distribution in bins of the e-peak charge, the dependence of the mean arrival time and the timing resolution on the e-peak size was studied. Specifically the mean arrival time is shown as a function of the e-peak pulse size for several drift voltages in Fig. \ref{fig:fig4} (left). The mean SAT dependence on the e-peak size found to follow a functional form of  Eq. (\ref{eq:powerlaw}).
\begin{equation}
g\left( x; \alpha, b, w\right) =\alpha+\frac{b}{x^w}
\label{eq:powerlaw}
\end{equation}
The power law parameter $w$, was found to be independent of the drift voltage settings but the constant term $a$ found to change with the drift voltage. The latter is easy to explain as the drift velocity depends on the drift voltage. However when the constant term has been subtracted all the points are following the same curve, independently of the drift voltage. Many tests \citep{noteb} have proven that the dependence of the arrival time on the e-peak size is not a systematic error due to the timing method. As shown in Fig. \ref{fig:fig4} (center), the average pulse shapes corresponding to different e-peak charge regions retain a constant shape but the higher pulses are arriving earlier, demonstrating that this ``time walk'' is related to a shift of the whole pulse for reasons that have nothing to do with the experimental procedure, but resulted by the pulse formation dynamics. In the right of Fig. \ref{fig:fig4}, the resolution's dependence on the size of the pulse (measured as the charge of the electron pulse) is shown. It is apparent that the resolution depends on the pulse size, but this dependence is the same for all drift voltages. Results presented until now are related with data taken with  600\,V voltage in the anode. Same dependences for SAT distributions and timing resolution are presented, where every colour corresponds to different anode voltage (left and central left of Fig. \ref{fig:fig5}). The fact is that the avalanches at the anode region do not affect the timing characteristics, but just amplify the signal. 
\begin{figure}
\begin{minipage}{.23\textwidth}
\centering
\includegraphics[width=.95\textwidth]{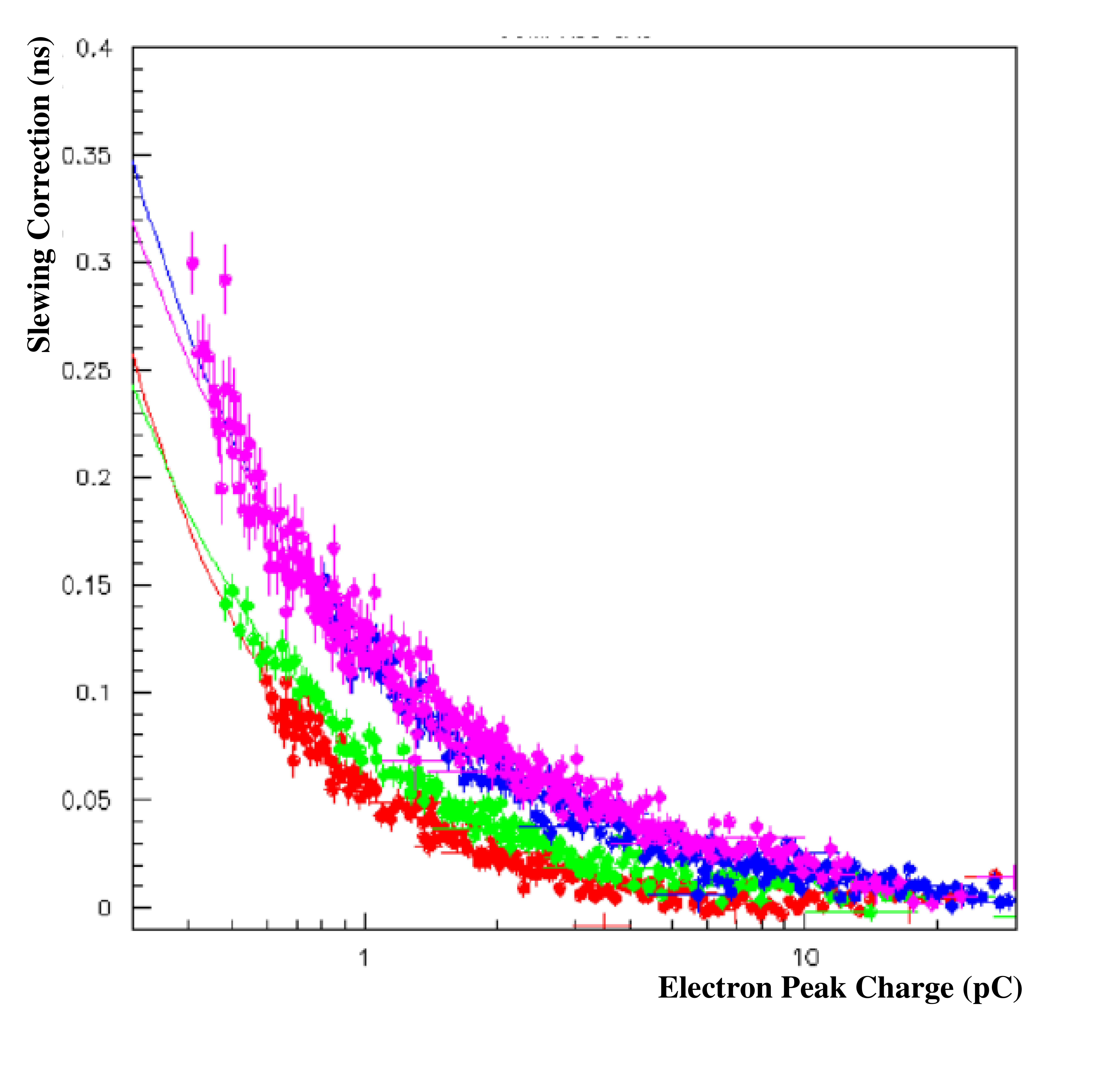}
\end{minipage}
\begin{minipage}{.23\textwidth}
\centering
\includegraphics[width=.95\textwidth]{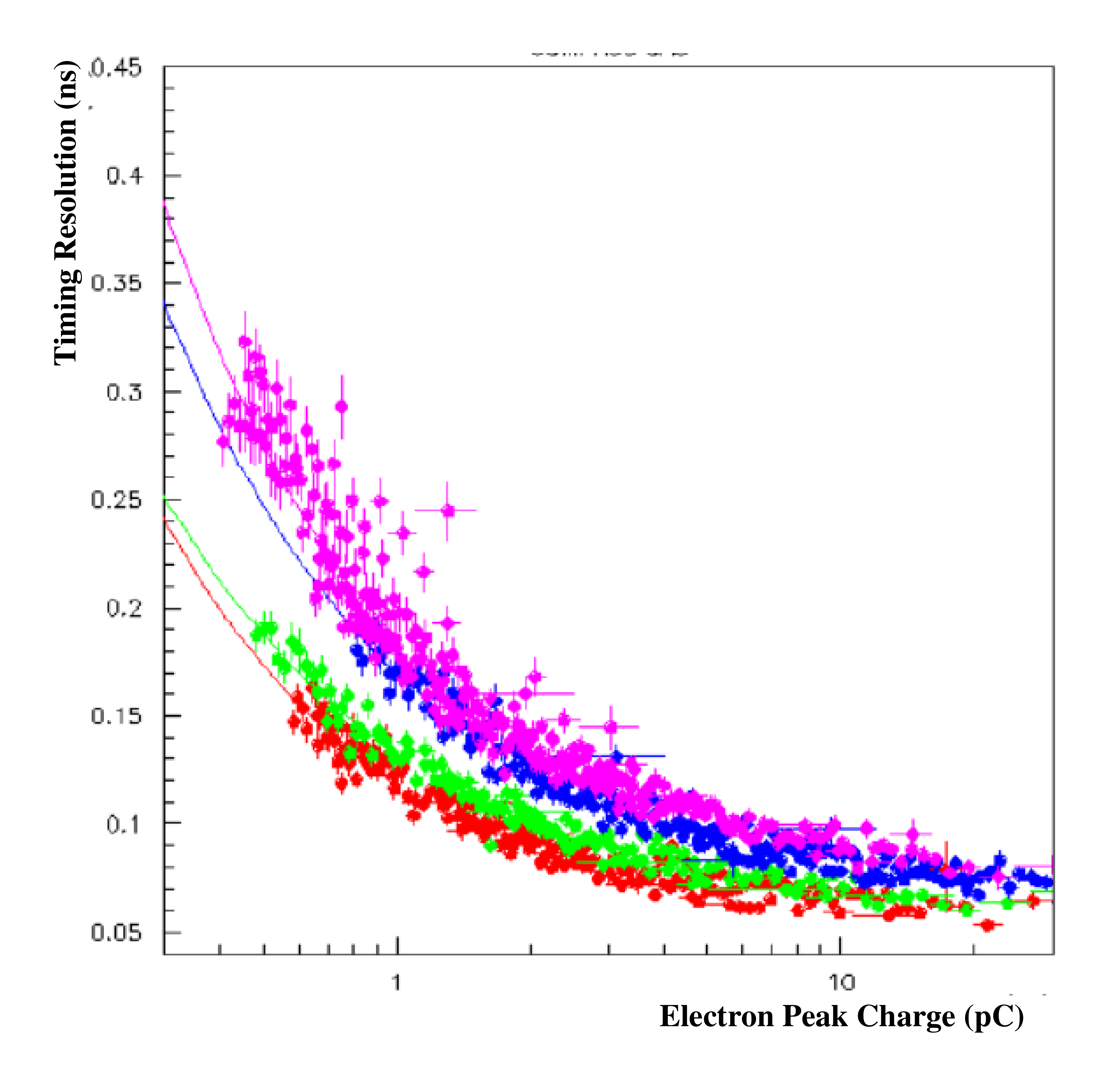}
\end{minipage}
\begin{minipage}{.25\textwidth}
\centering
\includegraphics[width=1.\textwidth]{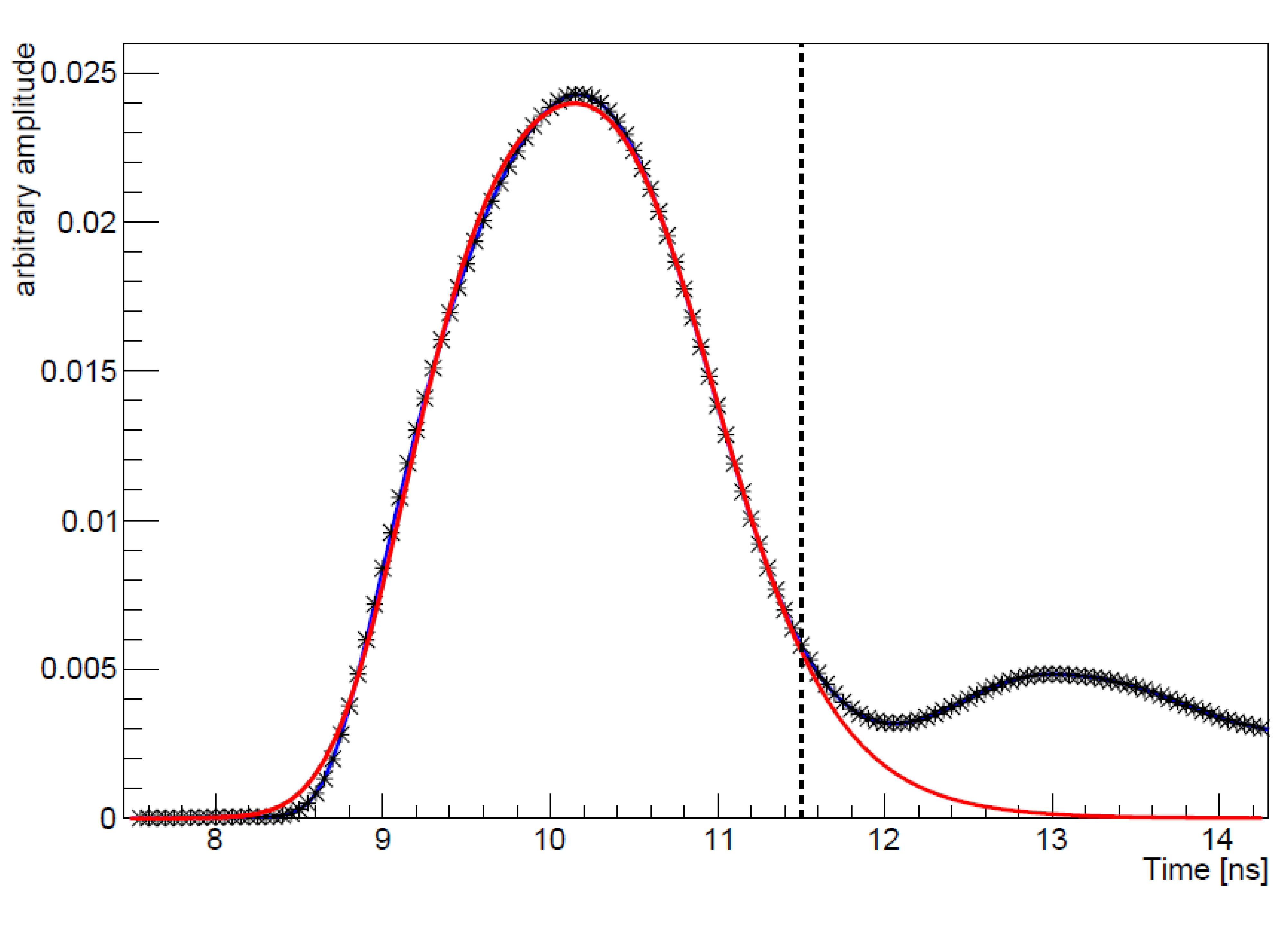}
\end{minipage}
\begin{minipage}{.25\textwidth}
\centering
\includegraphics[width=1.\textwidth]{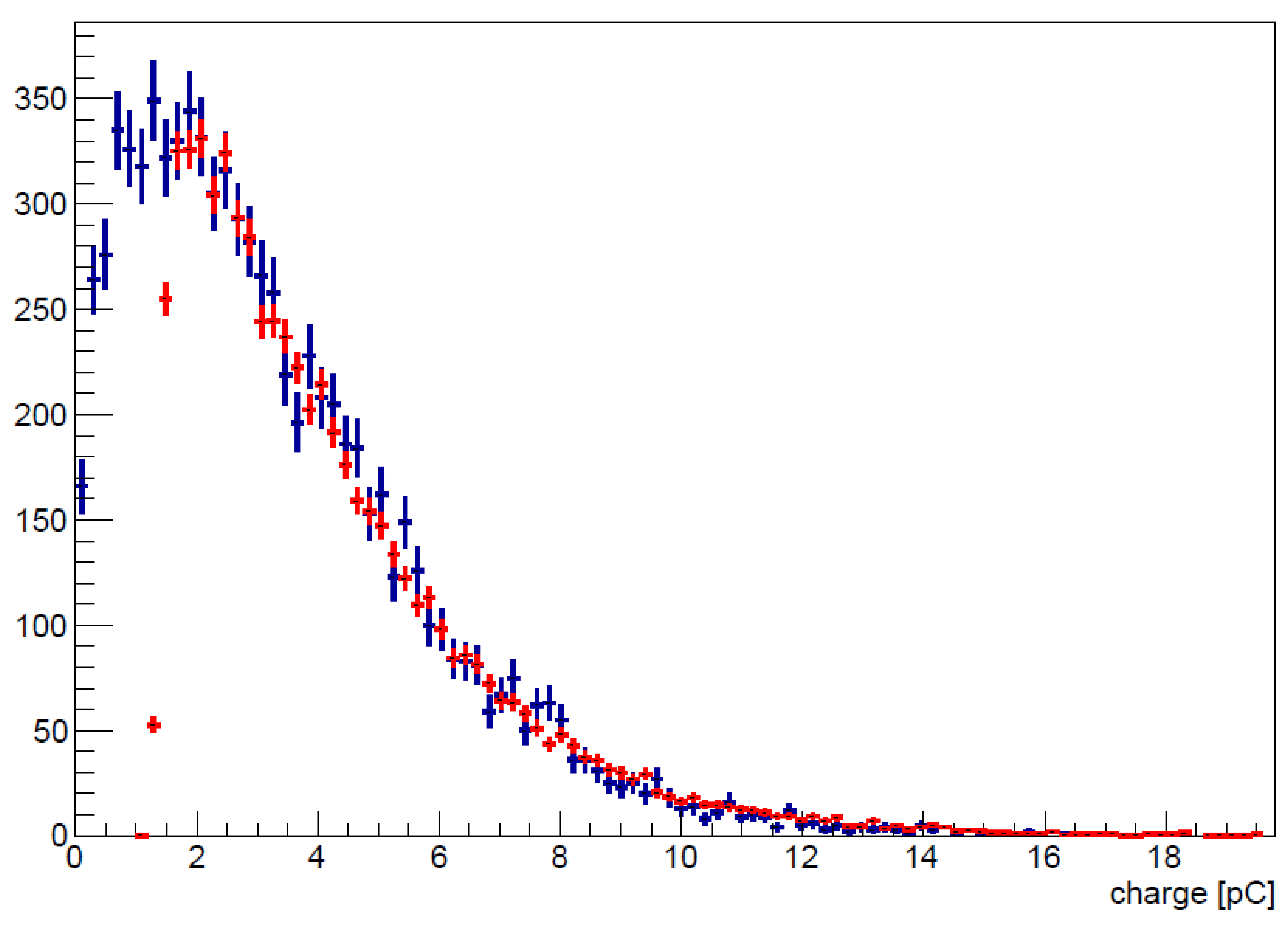}
\end{minipage}
\caption{The Mean SAT correction versus e-peak charge and time resolution versus e-peak charge with COMPASS gas mixture. Red corresponds to anode voltage 450\,V, green to 475\,V, blue to 650\,V and magenta to 525\,V (left and central left). The experimental average waveform (black points) for events with e-peak charge above 15\,pC. This region was chosen to prevent the SAT dependence affect the result. The fit result is shown in the red line of the plot. The dashed line corresponds to the right limit of the fit region (central right). Distribution of the simulated e-peak charge (blue), scaled with a factor G, such that the distribution matches the experimental one (red). Anode voltage is 450\,V and drift voltage is 400\,V. Scaling factor for this case is G = 27:8 (right).}
\label{fig:fig5}
\end{figure}
\section{SIMULATION STUDIES}
In order to investigate these finding a detailed simulation was developed, based on Garfield++ \citep{garfield}. In Ref. \citep{kostas, notec}, a technique which extracts the response of the electronics to a single amplification avalanche by comparing Garfield++ simulation with real data, is described in detail. Using the estimated response function for each electron arriving at the amplification region and assuming a linear response of the electronics, the simulation produces waveforms, including a 2.5\,mV RMS uncorrelated noise. As an example, in Fig. \ref{fig:fig5} (central right)  the average of real (black) and simulated (red) waveforms for e-peaks with charges around 15\,pC is presented. The chamber was operated with 450\,V and 425\,V anode and drift voltages, respectively. It should be noted that the prediction curve (in red) does not include the ion pulse. As a by-product of comparing distributions of simulated and  real data, the Penning transfer rate of the COMPASS gas filling was also estimated to be around 50\%.
\begin{figure}
\centering
\begin{minipage}{.30\textwidth}
\centering
\includegraphics[width=0.95\textwidth]{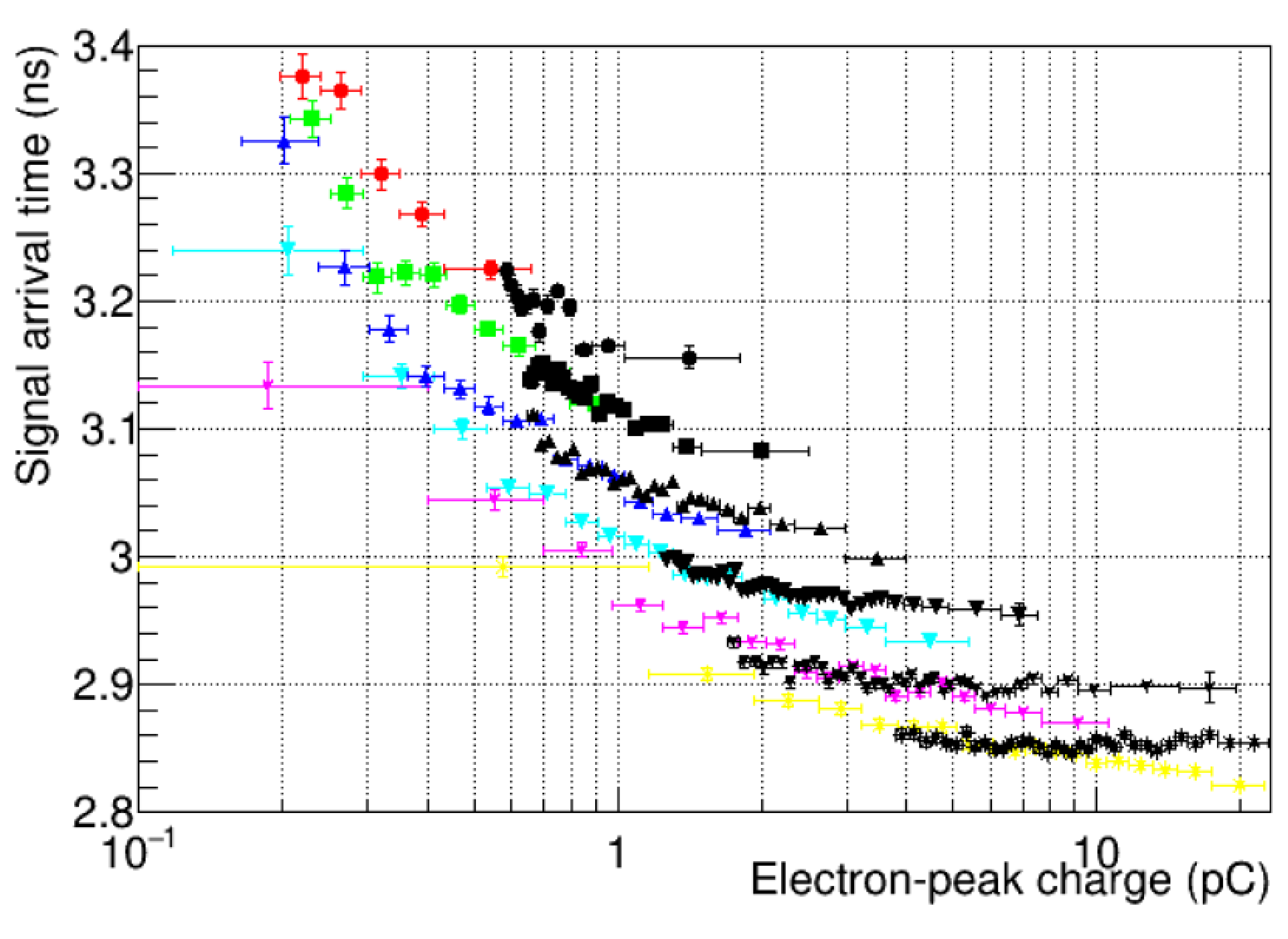}
\end{minipage}
\begin{minipage}{.30\textwidth}
\centering
\includegraphics[width=0.95\textwidth]{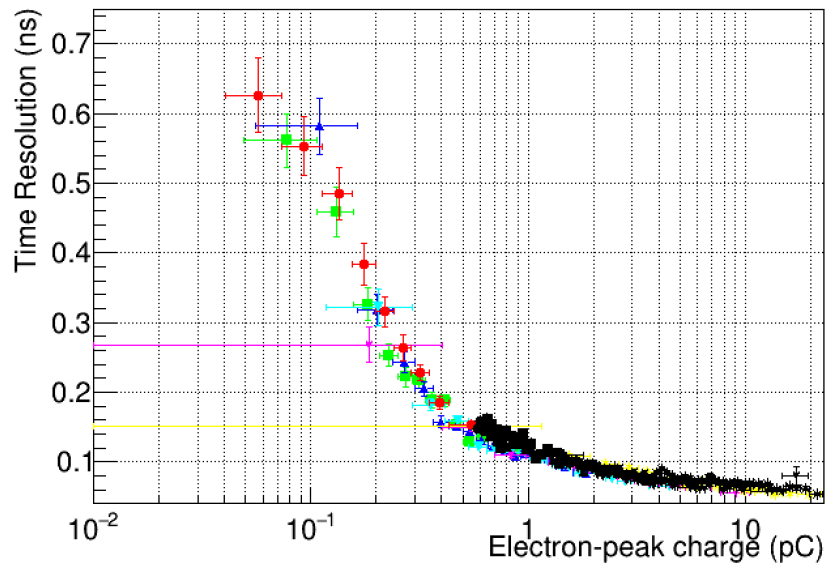}
\end{minipage}
\begin{minipage}{.30\textwidth}
\centering
\includegraphics[width=0.83\textwidth]{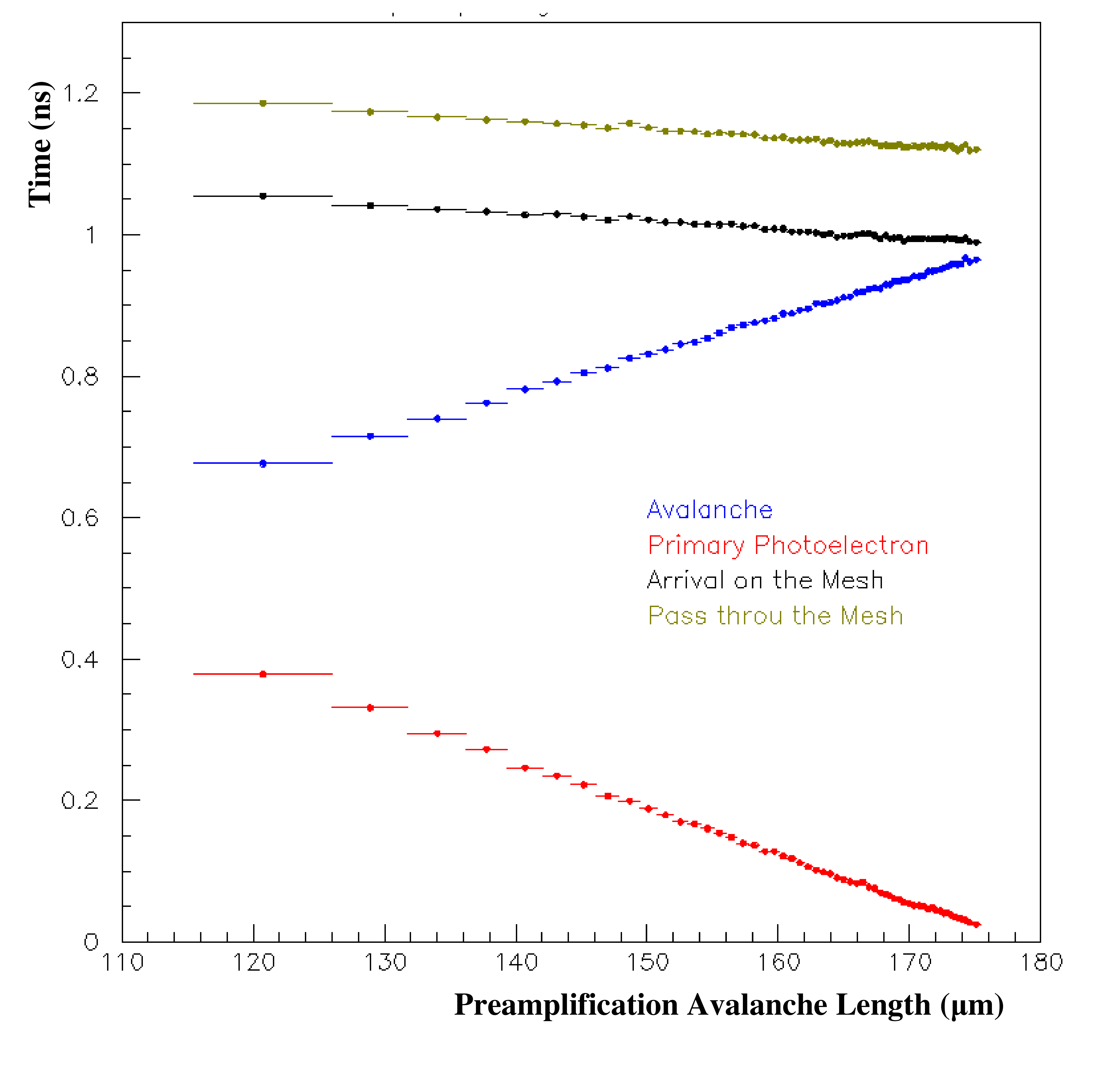}
\end{minipage}
\caption{Mean SAT as a function of the e-peak charge (left). Time resolution as a function of the e-peak charge (center). In both figures, black points correspond to experimental data while coloured correspond to simulated points with an anode voltage of 450\,V and for drift voltages of (red) 300\,V, (green) 325\,V, (blue) 350\,V, (cyan) 375\,V, (magenta) 400\,V and (yellow) 425\,V. Dependence of the mean transmission times with respect to the length of the avalanche (right).}
\label{fig:fig6}
\end{figure}
An other example demonstrating the successful simulation of the PICOSEC-MM response is shown on Fig. \ref{fig:fig5} (right), where the e-peak charge distribution for real data and simulated pulses are compared. The simulated pulses were digitised and analysed exactly the same way as the real waveforms. The resolution and the SAT were determined in bins of the e-peak charge. The dependence of the timing characteristics of the simulated waveform on the e-peak size is shown in left and center Fig. \ref{fig:fig6} in comparison to the respecting results found in the real data. The agreement between the experimental data and the simulation is apparent.
\begin{figure}
\centering
\begin{minipage}{.27\textwidth}
\centering
\includegraphics[width=1.\textwidth]{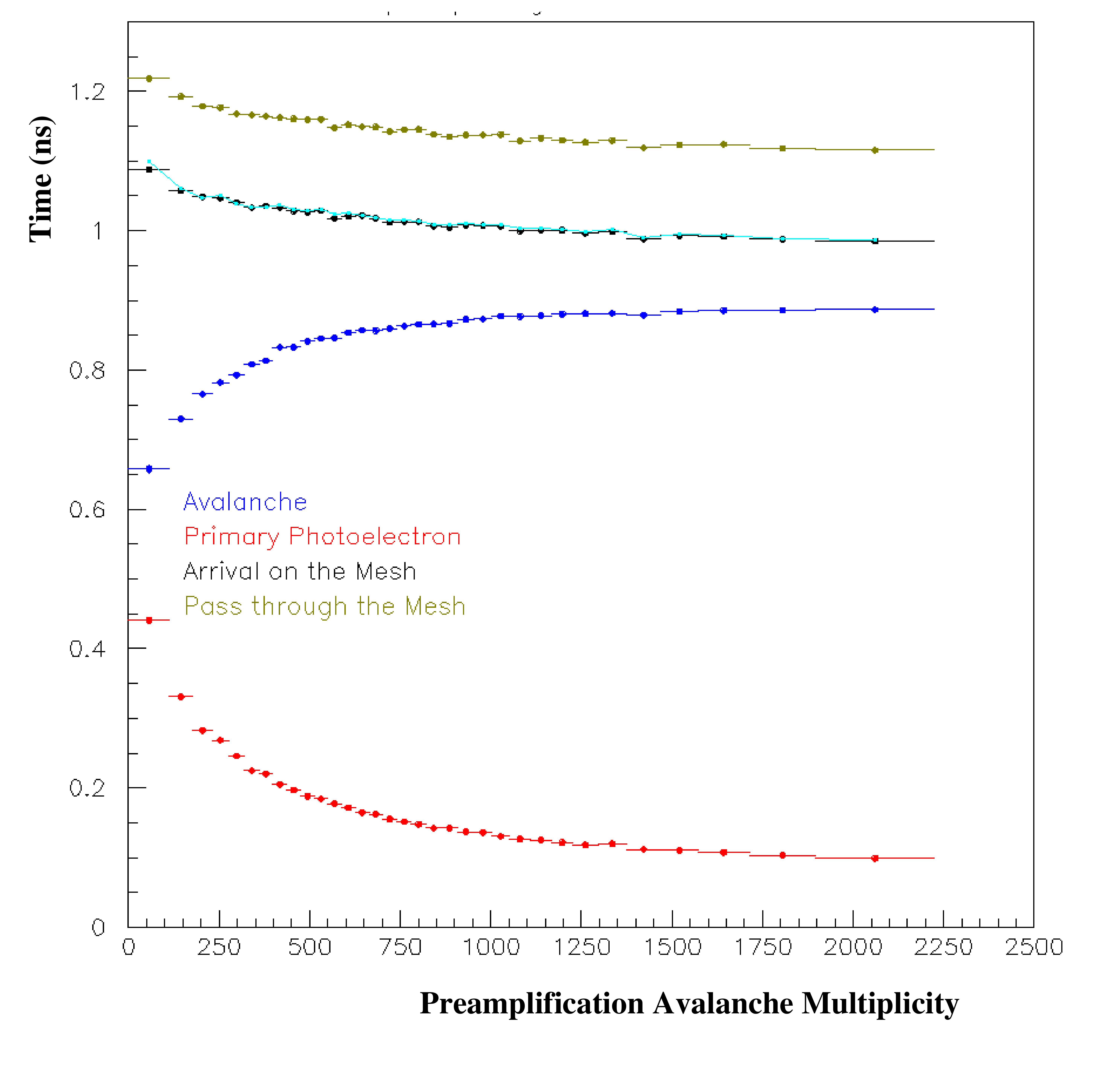}
\end{minipage}
\begin{minipage}{.27\textwidth}
\centering
\includegraphics[width=1.\textwidth]{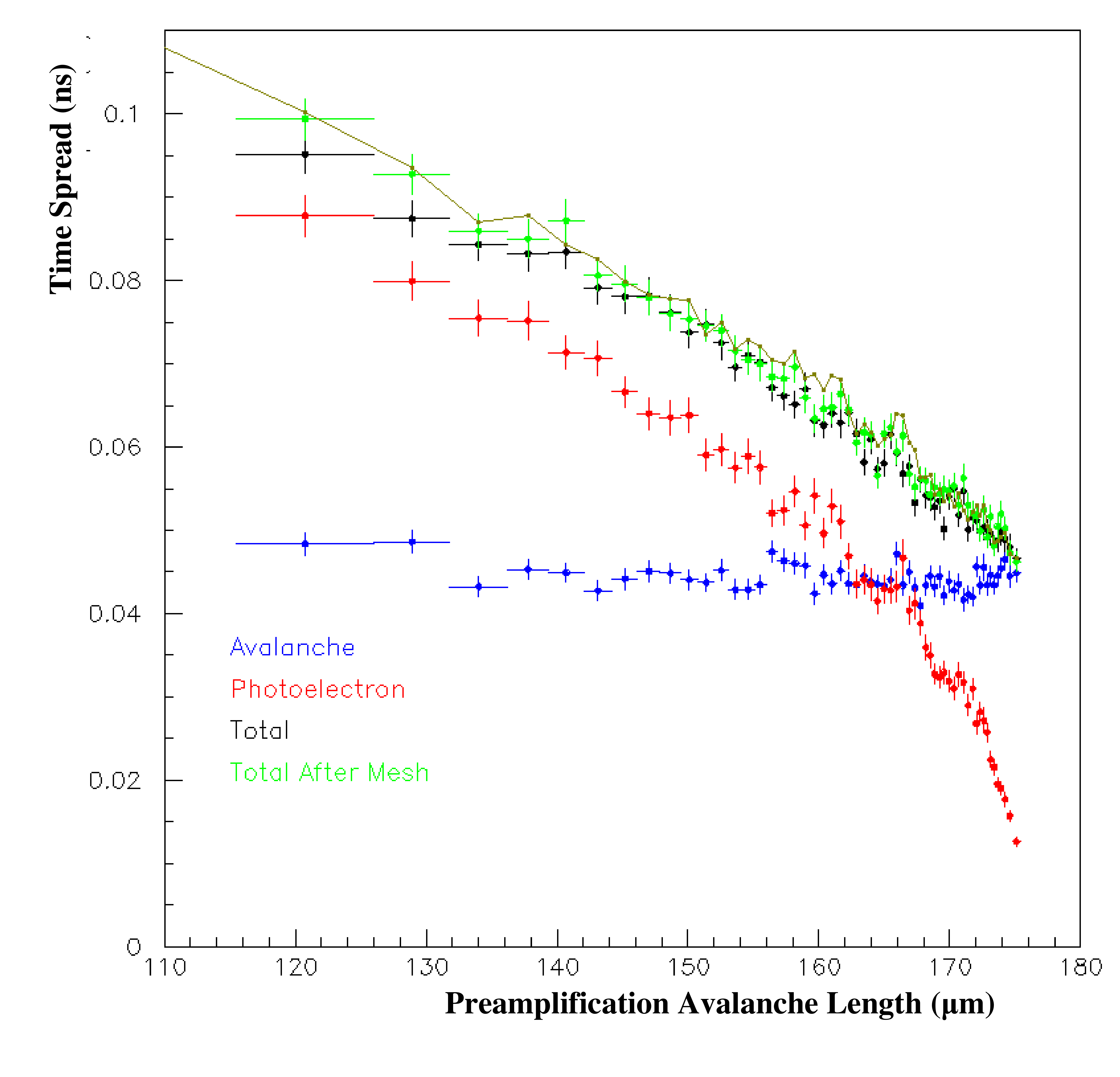}
\end{minipage}
\begin{minipage}{.27\textwidth}
\centering
\includegraphics[width=1.\textwidth]{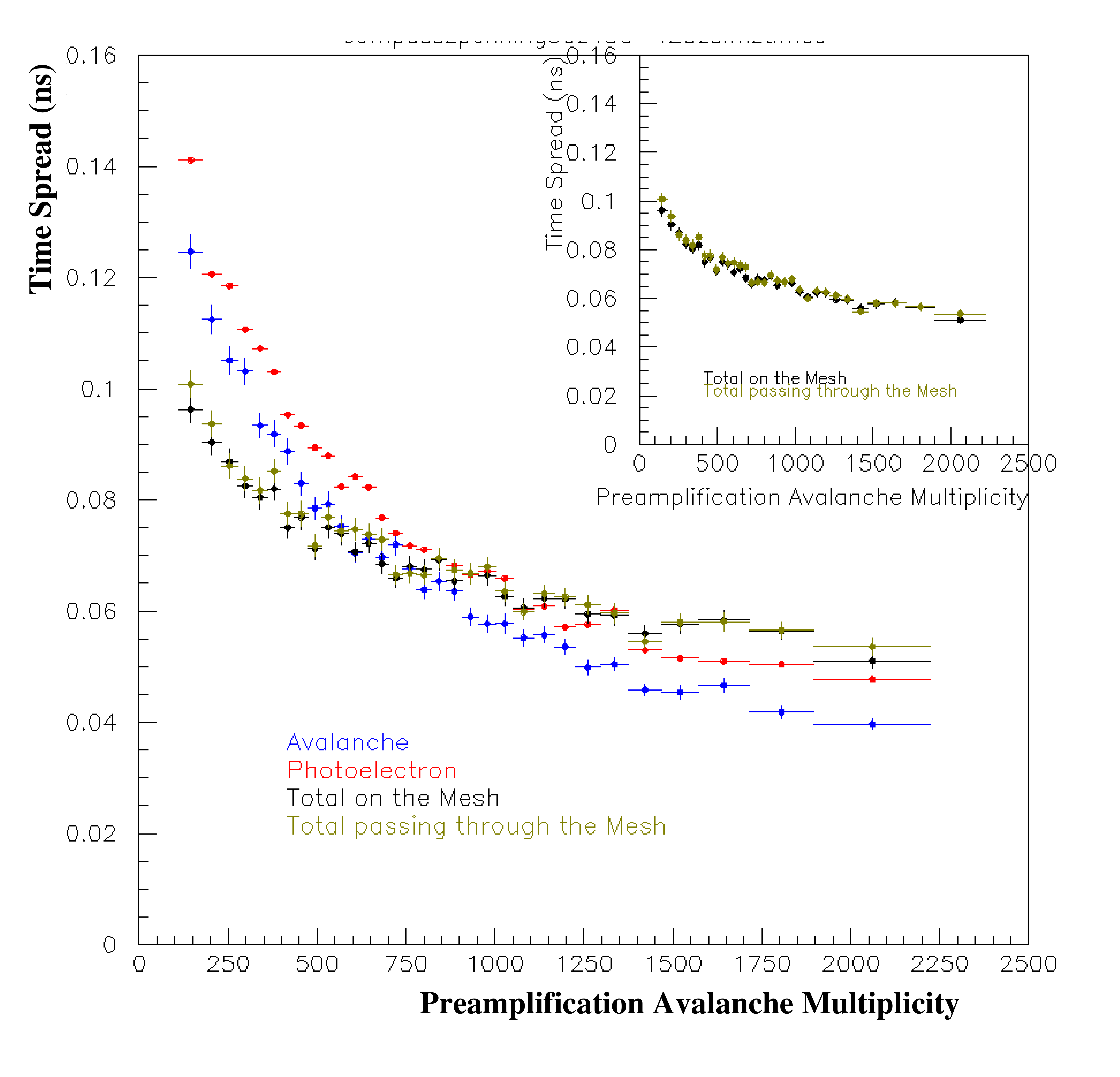}
\end{minipage}
\caption{Dependence of the mean transmission times with respect to the number of preamplification avalanche electrons (left). Dependence of the spread of the transmission times with respect to the length of the avalanche (center). Spread of the transmission times depending on the number of preamplification avalanche electrons (right).}
\label{fig:fig7}
\end{figure}
\section{PHENOMENOLOGICAL MODELLING}
Having established an agreement between Garfield++ prediction and the data, the next step was to identify the  microscopic physical parameters which correspond to the experimentally observed SAT. In Ref. \citep{kostas} was found that the microscopic parameter which corresponds to the SAT is the average of the times of those of the preamplification electrons passing through the mesh when they pass the mesh. This average time was found to have the same statistical properties as SAT, i.e. the same mean and RMS values in every bin of the e-peak charge. The statistical properties of this microscopic variable are determined by the transmission time of the photoelectron from its emission to the first ionisation point and on the effective drift time of the avalanche. The dependence of the mean transmission times with respect to the length of the avalanche is shown in Fig. \ref{fig:fig6} (right). Both the photoelectron (red) and the avalanche (blue) mean transmission times have linear relation with the length of the avalanche, whose sum (black) is not constant but also depends linearly on the avalanche length. The dependence of the mean time when passing through the mesh (green) differs only by a constant value. This dependence hints that the photoelectron and the avalanche drift with different velocities. Fig. \ref{fig:fig7} (left) shows (with the same colour code) the dependence of the above mean transmission times with respect to the, experimentally observable, number of preamplification avalanche electrons. This shows that there is  a ``time walk'' effect emerging in the microscopic simulation with Garfield++. The central plot of Fig. \ref{fig:fig7} presents the dependence of the spread of the transmission times with respect to the length of the avalanche, with the same color code. The spread of the photoelectron's transmission time increases with larger drift paths, while the spread of the avalanche's transmission time is saturated at a constant value. Therefore, the sooner the photoelectron ionizes for the first time, the better the time resolution. Fig. \ref{fig:fig7} (right) shows the spread of the transmission times depending on the number of preamplification avalanche electrons, with the same color code. Notice that the transmission time spreads of the photoelectron and the avalanche are larger than the total time spread. This is due to the fact that the photoelectron and avalanche transmission times are heavily correlated. 
\section{ACKNOWLEDGEMENTS}
We  acknowledge  the  financial  support  of  the  RD51  collaboration,  in  the  framework  of  RD51  common  projects,  the Cross-Disciplinary Program on Instrumentation and Detection of  CEA,  the  French  Alternative  Energies  and  Atomic  Energy  Commission;  and  the  Fundamental  Research  Funds  for the  Central  Universities of China. J.  Bortfeldt  acknowledges the support from  the  COFUND-FP-CERN-2014 program  (grant  number  665779). M. Gallinaro  acknowledges the  support  from  the  Fundac\={a}o  para  a Ci\^{e}ncia e a Tecnologia (FCT), Portugal (grants IF/00410/2012 and CERN/FIS-PAR/0006/2017).  D. Gonz\'{a}lez-D\'{i}az acknowledges  the  support  from  MINECO  (Spain)  under  the  Ramon y  Cajal  program  (contract  RYC-2015-18820).   F.J.  Iguaz  acknowledges  the  support  from  the  Enhanced  Eurotalents  program (PCOFUND-GA-2013-600382).  S. White acknowledges partial support through the US CMS program under DOE contract No. DE-AC02-07CH11359.
\bibliographystyle{unsrt}
\bibliography{pico_bib}
\end{document}